# COVID-19 publications: Database coverage, citations, readers, tweets, news, Facebook walls, Reddit posts

Kayvan Kousha, Statistical Cybermetrics Research Group, University of Wolverhampton, Wulfruna Street, Wolverhampton WV1 1LY, UK. ORCID: 0000-0003-4827-971X
Mike Thelwall, Statistical Cybermetrics Research Group, University of Wolverhampton, Wulfruna Street, Wolverhampton WV1 1LY, UK. ORCID:0000-0001-6065-205X

The COVID-19 pandemic requires a fast response from researchers to help address biological, medical and public health issues to minimize its impact. In this rapidly evolving context, scholars, professionals and the public may need to quickly identify important new studies. In response, this paper assesses the coverage of scholarly databases and impact indicators during 21 March to 18 April 2020. The results confirm a rapid increase in the volume of research, which particularly accessible through Google Scholar and Dimensions, and less through Scopus, the Web of Science, PubMed. A few COVID-19 papers from the 21,395 in Dimensions were already highly cited, with substantial news and social media attention. For this topic, in contrast to previous studies, there seems to be a high degree of convergence between articles shared in the social web and citation counts, at least in the short term. In particular, articles that are extensively tweeted on the day first indexed are likely to be highly read and relatively highly cited three weeks later. Researchers needing wide scope literature searches (rather than health focused PubMed or medRxiv searches) should start with Google Scholar or Dimensions and can use tweet and Mendeley reader counts as indicators of likely importance.

## Introduction

The international scientific effort to mitigate COVID-19 is unprecedented in scale and rapidity. For instance, PubMed added related publications daily between 1st January to 18th of April 2020[1], with a peak of over 300 in a single day. This effort is in response to the lethality and rapid spread of the disease, as well as the major economic and social consequences of COVID-19 lockdowns. As part of the response, researchers, professionals and the public will need to consult the scientific literature for the latest findings. Whilst this is normal for science, standard literature search methods may be ineffective in a rapid publishing environment. Traditional citation indexes may not be fast enough, especially given that they do not index most preprints, and citation counts may not help point to important studies. The more inclusive online citation indexes of sites like Google Scholar and Dimensions.ai seem like suitable alternatives since they index both the traditional scholarly literature and documents not published in journals, including preprints (Herzog, Hook, & Konkiel, 2020; Kousha, & Thelwall, 2019a). Whilst there are initiatives to help various communities with curated collections of COVID-19 documents, such as published biomedical documents from PubMed Central (PMC, 2020), preprints from medRxiv and bioRxiv (medRxiv, 2020), and a data mining collection (Allen Institute, 2020), none are complete. It is therefore important to assess the

---

[1] https://www.nlm.nih.gov/pubs/techbull/nd08/nd08_pm_new_date_field.html

COVID-19 coverage and growth of scholarly publication indexes, as well as the value of citation counts for new COVID-19 research.

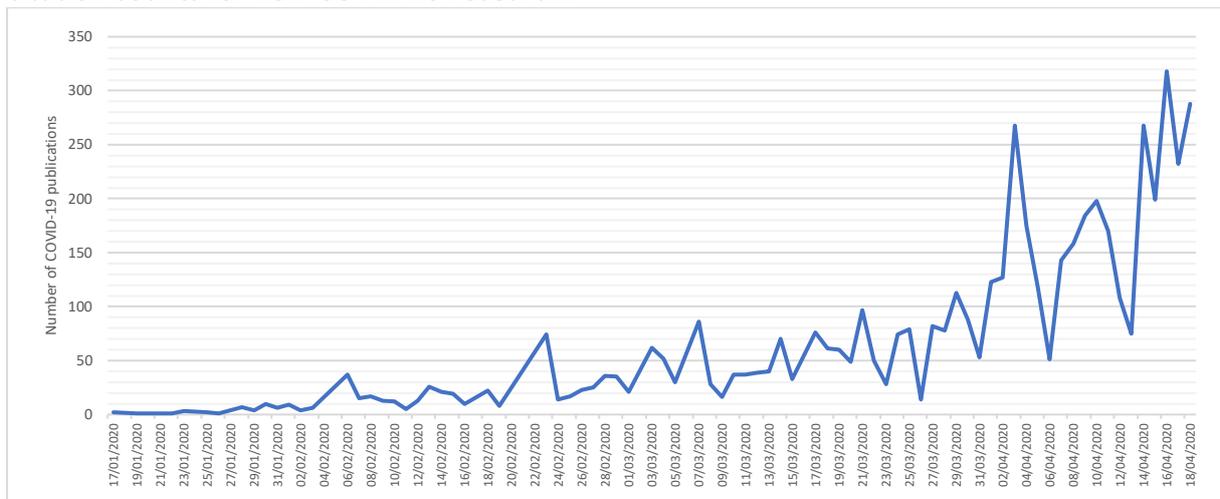

Figure 1. Daily additions of COIVD-19 publications to PubMed (17 Jan-18 April). Query used ((((((("COVID-19") OR "Novel coronavirus") OR "2019-nCoV") OR "SARS-CoV-2") OR "coronavirus 2") OR "Coronavirus disease 2019") OR Corona virus disease 2019) AND ("2019/12/01"[Date - Publication] : "3000"[Date - Publication]).

In parallel with scholarly needs for literature, the public, professionals and policy makers also need to access current COVID-19 research to inform their decision-making, such as whether to recommend wearing protective masks. This may be in addition to, or to clarify, World Health Organisation guidelines (WHO, 2020). They may therefore share relevant academic research in the social web (e.g., Merchant & Lurie, 2020), generating interest that may picked up by alternative indicators (altmetrics). Thus, altmetrics, may be useful in helping the public to identify the most relevant research or may help point researchers to topics considered important by the public. It would therefore be helpful to assess whether altmetrics can perform this role. In particular, since altmetrics can reflect both academic and non-academic interests (Mohammadi, Barahmand, & Thelwall, 2019; Mohammadi, Thelwall, Kwasny, & Holmes, 2018), it is not clear whether they will essentially be early indicators of citation impact or whether they reflect societal or other impacts for COVID-19. Altmetrics have already been shown useful to identify the spread of a misleading COVID-19 paper that was subsequently withdrawn (Ioannidis, 2020).

This paper addresses the above issues through a primarily descriptive analysis of the evolution of four online scholarly databases, and associated altmetrics, over four weeks in March-April 2020, when many countries were experiencing a lockdown. A previous study of 20 January to 12 April 2020 has shown continually increasing growth in the COVID-19 coverage of scholarly databases, with substantial variations between fields (Torres-Salinas, 2020). Individual highly cited or shared papers are also important to examine for qualitative insights into the types of research that are attracting substantial attention. The following research questions drive this paper.

- Which scholarly databases index the most COVID-19 publications (extending: Torres-Salinas, 2020)?
- Which COVID-19 documents have become highly cited or highly discussed?
- Do altmetrics and early citation counts reflect similar types of COVID-19 impact?

- Can any altmetrics serve as early indicators of future citation impact for COVID-19 documents?

## Background

The novel coronavirus 2019 (COVID-19) was first reported in Wuhan City, China in December. Quickly disseminating scientific results about COVID-19 is vital to allow quick action from successful clinical results (Song & Karako, 2020). The importance of scientific publishing to respond to infectious disease outbreaks has been emphasised by many bibliometric studies of previous cases (Rethlefsen & Livinski, 2013), such as SARS (Kostoff & Morse, 2011; Tian & Zheng, 2015), H7N9 influenza (Tian & Zheng, 2015), HIV/AIDS (Pouris & Pouris, 2011), Ebola (Pouris & Ho, 2016) and Zika (Delwiche, 2018).

One recent study using Dimensions, Scopus, WoS and the LitCovid (Chen, Allot, & Lu, 2020) curated list has investigated the daily growth of Covid-19 related publications in citation databases and digital libraries during 1st January to 7th of April, finding that Dimensions had best coverage (9,435 publications) compared to WoS (718) and Scopus (1,568). The weekly growth of PubMed was about 1,000 publications and the PubMed Central (1,398), medRxiv (989) and SSRN (608) repositories had best coverage of open access COVID-19 publications (Torres-Salinas, 2020). Google Scholar was not assessed and all evidence was extracted from Dimensions, so the counts for other repositories may not be complete.

### Dimensions citations

Dimensions.ai (Herzog, Hook, & Konkiel, 2020) is an online scholarly database that operates similarly to Google Scholar, in the sense of indexing documents using public information from the Web but has an Applications Programming Interface (API) that supports automatic downloading for all query matches. It indexes most documents in Scopus (Thelwall, 2018b), although not for all fields (Orduña-Malea & Delgado-López-Cózar, 2018). It seems to have substantial coverage of preprint servers, such as arXiv, and so probably has much larger coverage overall, especially for recently published papers. Its coverage seems to be higher than Scopus and the Web of Science (WoS), comparable to CrossRef but lower than Google Scholar and Microsoft Academic (Harzing, 2019). In line with this, citation counts for papers in Dimensions can be expected to be slightly higher than for Scopus and WoS but substantially higher for newer documents.

### Altmetrics: Mendeley Readers

Counts of readers from the social reference sharing site Mendeley form the most extensively researched and understood altmetric. A non-trivial minority of researchers (about 5%) used Mendeley by 2014 according to one survey, with disciplinary differences (Van Noorden, 2014). People typically register documents in Mendeley when they have read them or intend to read them (Mohammadi, Thelwall, & Kousha, 2016), so it is reasonable to regard Mendeley counts as an indicator of readership. According to self-reports in the site, users are predominantly academics and postgraduate students, with a few undergraduates, librarians and people in non-academic occupations (Mohammadi, Thelwall, Haustein, & Larivière, 2015). Thus, Mendeley is an indicator of predominantly academic readership, with an element of student readership.

A range of studies have investigated the relationship between Mendeley reader counts and citation counts, finding moderate or strong positive correlations (Costas, Zahedi,

& Wouters, 2015). Correlations between mature citation counts and Mendeley reader counts are strong and positive in almost all narrow fields in Scopus (Thelwall, 2017a), supporting their use as a citation impact type of indicator. Whilst the two types of data seem to be close to interchangeable for sets of mature articles (although they can differ sharply for individual education-oriented papers: Thelwall, 2017c), the advantage of Mendeley reader counts is that they appear and are useful a year before citation counts (Thelwall, 2017b). They may even be common enough to be used for scientometric purposes by the publication month of the publishing journal. Moreover, since early Mendeley reader counts correlate positively with later citation counts (Thelwall, 2018a), Mendeley reader counts are early academic impact indicators. They should therefore be a better academic impact indicator than citation counts for fast moving issues, such as COVID-19.

## Altmetrics: Tweeters, Facebook Walls

Twitter is a widely recognized source of altmetrics. More articles have non-zero tweet counts than non-zero scores on any other altmetric, other than Mendeley (Thelwall, Haustein, Larivière, & Sugimoto, 2013). As a news-oriented social media platform, articles can expect to get a substantial proportion of their tweets in the week of publication, so Twitter is visible long before citations. Nevertheless, tweeter counts (counting the number of tweeters rather than the number of tweets) are problematic to interpret. Whilst about half of people that tweet academic research are not academics (Mohammadi, Thelwall, Kwasny, & Holmes, 2018), tweets typically contain just article titles or brief summaries (Thelwall, Tsou, Weingart, Holmberg, & Haustein, 2013), serving as publicity rather than evidence of impact. Together with often close to zero correlations with citation counts (Costas, Zahedi, & Wouters, 2015; Haustein, Larivière, Thelwall, Amyot, & Peters, 2014; Thelwall, Haustein, Larivière, & Sugimoto, 2013), there is insufficient evidence to claim that tweeter counts are indicators of either academic or societal impact. Nevertheless, they may have some value for health-related research, where there is more public interest in academic research (Haustein, Larivière, Thelwall, Amyot, & Peters, 2014; Mohammadi, Gregory, Thelwall, Barahmand, 2020).

Facebook wall posts function like tweeter counts except that they are rarer (Costas, Zahedi, & Wouters, 2015; Thelwall, Haustein, Larivière, & Sugimoto, 2013). Since most of Facebook is private and Altmetric.com obtains its Facebook Wall counts only from public pages, this altmetric probably reflects a tiny fraction of all Facebook posts and may be oriented to organizational uses of Facebook (including journals) rather than typical users, and few posts are directly from academics (Mohammadi, Barahmand, & Thelwall, 2019).

## Altmetrics: News and Reddit

Altmetric.com harvests citations from online free news websites and the news-oriented site Reddit. Altmetrics from both are relatively rare and have very low correlations with citation counts (Costas, Zahedi, & Wouters, 2015; Thelwall, Haustein, Larivière, & Sugimoto, 2013). Nevertheless, health-related topics are newsworthy (Clark & Illman, 2006; Kousha, & Thelwall, 2019b), including for infectious diseases (e.g., SARS: Lewison, 2008) and so they may be useful for COVID-19.

# Methods

The research design is in three parts. First, to assess the relative coverage of scholarly databases, a range were queried daily from 21 March 2020 to record the number of COVID-19 documents indexed. Second, lists of documents matching a set of COVID-19 queries were downloaded from Dimensions.ai and altmetrics for these were gathered from Mendeley (Gunn, 2014) and Altmetric.com (Adie & Roe, 2013; Robinson-García, Torres-Salinas, Zahedi, & Costas, 2014) daily and the individual scores and documents compared. Third, a March-24 dataset was created to track a set of documents indexed on the same day.

## *Scholarly database indexing of COVID-19 publications*

To assess the indexing of COVID-19-related publications, the two mainstream scholarly databases, Scopus and Web of Science (WoS) were queried as well as a range of other academic sources that may index relevant documents. After testing, the core queries used to identify relevant documents were as follows. These are not comprehensive but are high precision, unless stated, and should include the most recent research focusing on the issue, assuming that it includes the current official disease description.

Table 1. COVID-19 queries for a range of scholarly sources.

| Source | Query | Scope/Year | Comments |
|---|---|---|---|
| Google Scholar | "COVID-19" | All/2019-2020 | OR does not work. May be inaccurate. |
| Dimensions | "COVID-19" OR "Novel coronavirus" OR "2019-nCoV" OR "SARS-CoV-2" OR "coronavirus 2" OR "Coronavirus disease 2019" OR "Corona virus disease 2019" | All fields and publication types/2019-2020 | |
| PubMed | ((((((("COVID-19") OR "Novel coronavirus") OR "2019-nCoV") OR "SARS-CoV-2") OR "coronavirus 2") OR "Coronavirus disease 2019") OR Corona virus disease 2019) AND ("2019/12/01"[Date - Publication] : "3000"[Date - Publication]) | All fields and publication types from Dec 2019 | |
| Mendeley | "COVID-19" | All/not assigned | OR does not work |
| medRxiv and bioRxiv | Self-reported repository statistics for self-curated collection. | Full text papers | Repository statistics for COVID-19 SARS-CoV-2 preprints from medRxiv and bioRxiv[2]. |
| Scopus | (ALL ( "COVID-19" ) OR ALL ("Novel coronavirus") OR ALL ("2019-nCoV") OR ALL ("SARS-CoV-2") OR ALL ("coronavirus 2") OR ALL ("Coronavirus disease 2019") OR ALL ("Corona virus disease 2019" ) ) AND PUBYEAR = 2020 OR PUBDATETXT (december 2019) | All fields and publication types/2019-2020 | |
| WoS Core Collection | TOPIC=("COVID-19" OR "Novel coronavirus" OR "2019-nCoV" OR "SARS-CoV-2" OR "coronavirus 2" OR "Coronavirus disease 2019" OR "Corona virus disease 2019") | All fields and publication types /2019-2020 | Including Conference Proceedings Citation Index. |
| PMC | (((((((("COVID-19") OR "Novel coronavirus") OR "2019-nCoV") OR "SARS-CoV-2") OR "coronavirus 2") OR "Coronavirus disease 2019") OR "Corona virus disease 2019") AND ("2019/12/01"[Publication Date] : "3000"[Publication Date]) | Full text publications from Dec 2019 | |
| Clinical Trials.gov | COVID OR "SARS-CoV-2" OR "2019-nCoV" | | Query predefined statistics[3]. |

*Document and altmetric comparison datasets*

Initial testing suggested that Dimensions and Google Scholar had the largest coverage of COVID-19 documents. Since Google Scholar does not have an API and the number of matches exceeds its 1000 limit per query, it was not possible to extract Google Scholar's set of matching documents. In contrast, Dimensions.ai has an API allowing complete sets of matching document records to be downloaded and so it was chosen as the base source of COVID-19 documents. It was checked daily with the following set of queries in the Dimensions API (Applications Programming Interface), designed to match publications about COVID-19 using various related names. These queries are all designed to be precise but there were still a few false matches. All queries ended in, "return publications [basics + extras]"

- search publications for "COVID-19" where year >= 2019
- search publications for "Novel coronavirus" where year >= 2019
- search publications for "2019-nCoV" where year >= 2019
- search publications for "SARS-CoV-2" where year >= 2019
- search publications for "coronavirus 2" where year >= 2019
- search publications for "Coronavirus disease 2019" where year >= 2019
- search publications for "Corona virus disease 2019" where year >= 2019

The resulting 21,395 publications were mainly open access (53%; 75% for the March 24 set – see later) and predominantly from health-related specialties (Table 2).

Table 2. The top 10 Dimensions subject codes for the complete and March 24 datasets.

| FOR code | All | All | % | Mar-24 | Mar-24 | % |
|---|---|---|---|---|---|---|
| 1117 Public Health and Health Services | 3072 | 2762.9 | 13% | 78 | 73.3 | 21% |
| 1108 Medical Microbiology | 2773 | 2240.8 | 10% | 32 | 27.2 | 8% |
| 1103 Clinical Sciences | 2159 | 1860.9 | 9% | 32 | 28.8 | 8% |
| 0601 Biochemistry and Cell Biology | 1192 | 946.3 | 4% | 14 | 11.3 | 3% |
| 1107 Immunology | 1096 | 873.4 | 4% | 5 | 2.8 | 1% |
| 0604 Genetics | 803 | 642.7 | 3% | 4 | 2.8 | 1% |
| 1102 Cardiorespiratory Medicine & Haematology | 459 | 384.4 | 2% | 7 | 6.5 | 2% |
| 0801 Artificial Intelligence and Image Processing | 383 | 336.0 | 2% | 6 | 6.0 | 2% |
| 1109 Neurosciences | 316 | 257.9 | 1% | 2 | 1.3 | 0% |
| 0605 Microbiology | 364 | 224.3 | 1% | 0 | 0.0 | 0% |

The datasets analysed include substantial numbers of papers from preprint planforms, including medRxiv, SSRN, arXiv, bioRxiv, ChemRxiv and Research Square (Table 3, as in: Torres-Salinas, 2020) as well as books and more traditional journals (Table 3).

Table 3. The top 10 journals, as recorded in Dimensions, for the complete and March 24 datasets.

| Journal | All | % | Mar-24 | % | Comment |
|---|---|---|---|---|---|
| [None] | 2932 | 14% | 13 | 4% | Books, book chapters, theses |
| medRxiv | 1234 | 6% | 30 | 9% | Health sciences preprints |
| SSRN Electronic Journal | 855 | 4% | 0 | 0% | Social science preprints |
| arXiv | 389 | 2% | 16 | 5% | Physics/computing preprints |
| bioRxiv | 358 | 2% | 1 | 0% | Biological sciences preprints |
| Research Square | 341 | 2% | 13 | 4% | Preprint platform |
| The BMJ | 262 | 1% | 9 | 3% | Core medical journal |
| ChemRxiv | 210 | 1% | 8 | 2% | Chemistry preprints |
| Viruses | 196 | 1% | 1 | 0% | MDPI open access journal |
| Journal of Medical Virology | 176 | 1% | 4 | 1% | Wiley journal |

Although most documents were classified as Articles by Dimensions, this type includes medRxiv preprints and diverse types of document published in journals, such as notes, short communications, editorials and commentaries (Table 4). Since many editorials seemed to discuss the impact of COVID-19 on the journal or field, this added less citable documents to the Article class. The surprising number of books and book chapters (13% overall) seems to be primarily due to pre-COVID-19 discussions about Coronaviruses, matching the query "Coronavirus 2". The low number of conference proceedings may be due to conference cancellations, or the inability of most conferences to respond to the COVID-19 timescale.

Since the Dimensions type Article includes documents that would not be classed as standard journal articles in scientometric analyses, the 295 Dimensions "Articles" from March 24 were visited to classify them by type. Only 106 of these seemed to be standard journal articles. The rest were mainly editorials, letters (called letters, letters to the editor, or correspondence; one Letter was classed as an article) or news stories. In some cases, documents were called "article" by the publishing journal but were clearly news stories published in a news-focused magazine/journal. The reduced set of 106 journal articles from March 24, 2020 was used for follow-up correlation tests.

Table 4. The top 10 document types, as recorded in Dimensions, for the complete and March 24 datasets.

| Type | All | % | Mar-24 | % | Comments |
|---|---|---|---|---|---|
| Article | 16330 | 76% | 295 | 85% | Includes preprints from medRxiv, editorials, commentaries |
| Book | 832 | 4% | 4 | 1% | Matches more general "Coronavirus 2" research |
| Chapter | 1645 | 8% | 5 | 1% | Matches more general "Coronavirus 2" research |
| Preprint | 2236 | 10% | 43 | 12% | Includes arRxiv, Research Square, chemRxiv, JMIR Preprints, SSRN |
| Monograph | 166 | 1% | 2 | 1% | Matches more general "Coronavirus 2" research |
| Proceeding | 186 | 1% | 0 | 0% | Conference proceedings |
| Total | 21392 | 100% | 349 | 100% | |

After Webometric Analyst had downloaded a complete set of records each day, the Mendeley API was used to identify the number of Mendeley readers for each document, again using Webometric Analyst. It queries by DOI and by title/author/year and combines non-overlapping results for the most complete reader count. This follows best practice (Zahedi, Haustein, & Bowman, 2014). Webometric Analyst was also used to identify counts of citations in Twitter, Facebook, Reddit and online news outlets to these documents, as identified by DOI queries to Altmetric.com. This data provider seems to have the most comprehensive coverage of Twitter, the largest of the sources (Ortega, 2018). Twitter and Facebook are logical choices to investigate because they seem to be the social media sources that most cite academic research (Costas, Zahedi, & Wouters, 2015; Thelwall, Haustein, Larivière, & Sugimoto, 2013). Reddit and news may give a news perspective, although Reddit is a multipurpose site (Ovadia, 2015; Stoddard, 2015) and the news sources harvested by Altmetric.com presumably exclude some major paywalled press sources.

There were some gaps in the data collection due to documents not being returned by a query on one day when they had been returned on a previous day. This produced missing citation and altmetric scores, affecting the analysis. To avoid this issue, these missing values were replaced by approximate values by linear interpolation (when scores were available for previous and subsequent dates), linear extrapolation (when at least two previous but no subsequent scores were available), or constant values (when only one previous value was available).

## Analysis

The coverage of the different sources was evaluated by comparing (on a graph) the number of query matches over time. This is not a fair comparison because the queries are not equivalent, a researcher may use other queries, and the sources index with different levels of comprehensiveness. For example, a source that indexed full text of documents would get more and probably less relevant hits than a source indexing the title and abstract, even if they had the same coverage.

To assess the types of document generating the most impact for each source, the top 5 for each indicator was extracted to give a manageable set. A comparison of the relative ranks of these documents for the different indicators was used to guide the evaluation of the relative importance of the document characteristics, along with the document age (younger documents would tend to have lower scores in less rapidly evolving indicators). This focus on the highest scoring documents seems reasonable since they are likely to be the most influential or important, even though different trends may apply to more average documents.

To compare the average accumulation speed and scores of COVID-19 documents, a base set was chosen, consisting of documents first indexed in Dimensions on 24 March 2020. This was the date from the first week with the most new documents (excluding the first day). These documents form a set that are likely to have been published on or shortly before 24 March 2020. The altmetric and citation scores for this set were compared over time to assess their evolution and relative magnitude. Averages were calculated with geometric means (with a +1 offset: Fairclough, & Thelwall, 2015) rather than arithmetic means due to the highly skewed nature of citations (de Solla Price, 1976; Wallace, Larivière, & Gingras, 2009) and altmetrics (Thelwall & Wilson, 2016; Yu, Xu, Xiao, Hemminger, & Yang, 2017). The scores of this set were then compared using Spearman correlations to assess the extent to which they may reflect similar types of impact (Sud & Thelwall, 2014). Since altmetrics other than Mendeley tend to have very weak correlations with citation counts (Costas, Zahedi, & Wouters, 2015; Haustein, Larivière, Thelwall, Amyot, & Peters, 2014; Thelwall, Haustein, Larivière, & Sugimoto, 2013), high correlations are not expected. Field normalization was not used for either analysis because (a) the papers cover a relatively narrow topic (COVID-19) even though they span many subject areas and (b) it is impractical to field normalize the values because this would require daily updates of the whole of Dimensions, Altmetric.com and Mendeley for the calculations.

# Results

## *Coverage of scholarly databases*

Based on the estimated number of manual search results returned by the sources queried, it seems that Google Scholar has substantially wider coverage of COVID-19 publications than all other sources (Figure 2). The results for Google Scholar may well be substantially inflated by its web search component indexing advertisements or warnings in webpages alongside articles irrelevant to the disease, however, so these results are not robust. To illustrate the existence of these false matches, a search for "COVID-19" in Google Scholar with a date range specified as 1990-2000 (i.e., 20 years before the name was coined) on April 21, 2020 returned an estimated 5,010 matches[4]. Each incorrect Google Scholar match reported snippets not from the paper, such as, "PEDIATRICS COVID-19 COLLECTION We are fast-tracking and publishing the latest research and articles related to COVID-19 for free." Nevertheless, it seems likely that Google Scholar indexes at least as many documents as Dimensions.

Google Scholar and Dimensions index both publisher records and other online publications (preprint archives for Dimensions, wider web sources for Google Scholar) seem able to identify COVID-19 publications more quickly or more widely than the Web of Science and Scopus. This suggests that academics studying the area should consider them if more specialist databases, such as PubMed, are not adequate. This argument does not take into account the importance of the documents, however, and it is possible that the key publications are quickly peer reviewed, published and indexed by Scopus and WoS. The exact COVID-19 coverage of Google Scholar is difficult to assess because it is not possible to download and check all matches in the absence of a Google Scholar API to download large sets of publication records.

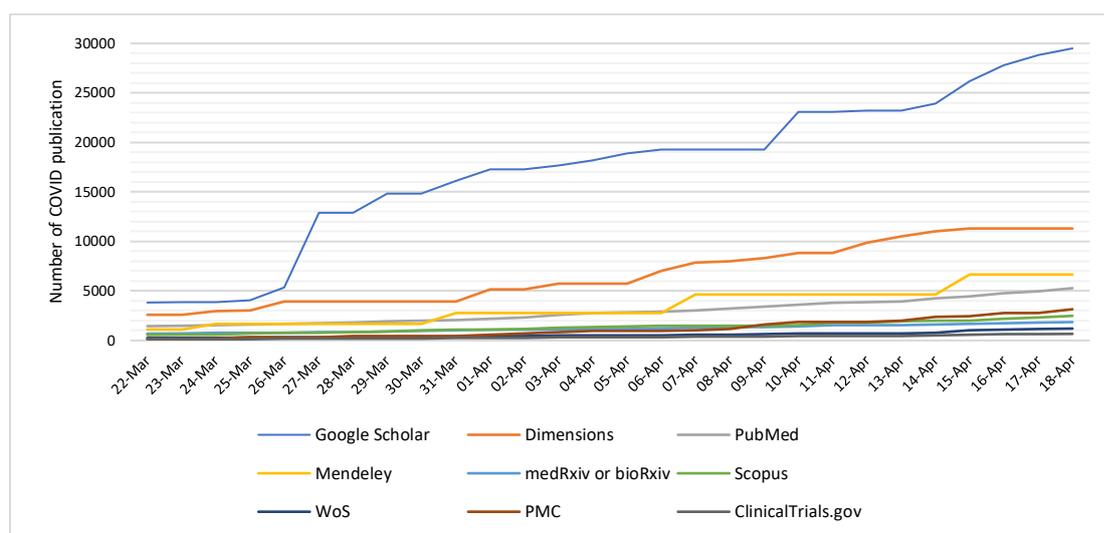

Figure 2. The daily number of hits for COVID-19 queries (see Table 1) from a range of scholarly sources (22 March – 18 April). The Google Scholar estimates include many false matches.

---

[4] https://scholar.google.co.uk/scholar?q=covid-19&hl=en&as_sdt=0%2C5&as_ylo=1990&as_yhi=2000

*Overlaps between Dimension, Scopus and Web of Science*

The extent of overlaps between the COVID-19 query results for Dimensions, Scopus and WoS were estimated on 19 April, 2020 to assess whether they were indexing the same publications. To obtain a relevant set of COVID-19 publications, only publications from 2019-2020 with the term "COVID" OR "coronavirus" OR "2019-nCoV" OR "SARS-CoV-2" OR "Corona" in their titles were selected. Publications with DOIs were matched between the three databases to assess the percentage overlap between them (Table 5).  Few of the Dimensions publications were also in Scopus (23.3%) or WoS (11.8%). Two fifths (40.4%) of the Scopus publications were in WoS  and four fifths (81.9%) of WoS publications were in Scopus. Google Scholar could not be compared without a comprehensive list of search matches.

Table 5. Overlaps for COVID-19 publications between Dimension, Scopus and Web of Science (18 April, 2020).

| | Total publications | Overlap % (No.) | | Non-overlapping % (No.) | |
|---|---|---|---|---|---|
| | | Scopus | WoS | Scopus | WoS |
| **Dimensions** | 8,642 | 23.3% (2,010) | 11.8% (1,017) | 76.7% (6,632) | 88.2% (7,624) |
| **Scopus** | 2,166 | - | 40.4% (874) | - | 59.6% (1,292) |
| **WoS** | 1,067 | 81.9% (874) | - | 18.1% (193) | - |

In terms of citations found by the three databases for the matching publications, Dimensions citation counts for all its matching COVID-19 publications were 4.9 and 2.8 times as numerous as WoS and Scopus, suggesting for the recently published or in-press articles, Dimensions had faster citation indexing than WoS and Scopus or from faster sources, such as preprint archives. This could be important when scholars want to consider early citation impact evidence for identifying relevant COVID-19 publications or for the impact assessment of published articles.

*Most cited papers.*

The documents with the most Mendeley readers and Dimensions citations tended to be similar and to provide primary clinical and epidemiological evidence about COVID-19 (Table 6). Shorter publication formats and analyses are more evident in the social web and news sources, representing a partially different type of document. The social web and news articles also seemed to give information that might be particularly useful as public health information for the vast majority of the planet's population that had not yet caught COVID-19 by 18 April 2020. These include studies on facemasks, the stability of the virus on surfaces, and pregnancy risks.

Table 6. Characteristics and ranks of COVID-19 papers in the top **five** for **D**imensions, **M**endeley, **T**witter, **F**acebook, and **N**ews, and their ranks in these sites on 18 April 2020. Citation and altmetric counts are in the figures below.

| Title | Journal* | Date | Type | D | M | T | F | N |
|---|---|---|---|---|---|---|---|---|
| Clinical features of patients infected with 2019 novel coronavirus in Wuhan, China | Lancet | 24/1/20 | Article | 1 | 1 | | | |
| A novel coronavirus from patients with pneumonia in China, 2019 | NEJM | 20/2/20 | Brief report | 2 | 4 | | 3 | |
| Early transmission dynamics in Wuhan, China, of Novel coronavirus-Infected pneumonia | NEJM | 26/3/20 | Article | 3 | 3 | | | |
| Epidemiological and clinical characteristics of 99 cases of 2019 novel coronavirus pneumonia in Wuhan, China: a descriptive study | Lancet | 30/1/20 | Article | 4 | | | | |
| Clinical characteristics of 138 hospitalized patients with 2019 novel Coronavirus-Infected pneumonia in Wuhan, China | JAMA | 7/2/20 | Original Investigation | 5 | | | | |
| Clinical characteristics of coronavirus disease 2019 in China | NEJM | 28/2/20 | Article | | 2 | 4 | | |
| Clinical course and risk factors for mortality of adult inpatients with COVID-19 in Wuhan, China: a retrospective cohort study | Lancet | 11/3/20 | Article | | 5 | | | 3 |
| The proximal origin of SARS-CoV-2 | Nature Medicine | 17/3/20 | Correspondence | | | 1 | 4 | |
| Treatment of 5 critically ill patients with COVID-19 with convalescent plasma | JAMA | 27/3/20 | Preliminary Comm. | | | 2 | | |
| Respiratory virus shedding in exhaled breath and efficacy of face masks | Nature Medicine | 3/4/20 | Brief Comm. | | | 3 | | |
| Aerosol and surface stability of SARS-CoV-2 as compared with SARS-CoV-1 | NEJM | 17/3/20 | Correspondence | | | 5 | | 1 |
| Coronavirus latest: CERN scientists join the COVID-19 fight. | Nature | 8/4/20 | News | | | | 1 | |
| Clinical characteristics and intrauterine vertical transmission potential of COVID-19 infection in nine pregnant women: a retrospective review of medical records | Lancet | 12/2/20 | Article | | | | 2 | |
| Characteristics of and important lessons from the coronavirus disease 2019 (COVID-19) outbreak in China | JAMA | 24/2/20 | View-point | | | | 5 | 4 |
| The incubation period of coronavirus disease 2019 (COVID-19) from publicly reported confirmed cases: estimation and application. | Annals of Internal Medicine | 10/3/20 | Original research | | | | | 2 |
| Severe outcomes among patients with coronavirus disease 2019 (COVID-19) - United States, February 12-March 16, 2020. | Morbidity Mortality Weekly Report | 18/3/20 | Report | | | | | 5 |

*NEJM: New England Journal of Medicine

None of the five documents most cited on Reddit were also in the top five for the other sources, although they seem to cover similar topics (Table 7). The paper about Malayan pangolins is the exception for not covering the primary characteristics of the disease or public health issues. This may be an artefact of the relatively low numbers of Reddit citations.

Table 7. Characteristics and ranks of COVID-19 papers in the top **five** for Reddit, and their ranks in these sites on 18 April 2020. There is no overlap with Table 1. Citation and altmetric counts are in the figures below.

| Title | Journal | Date | Type | R |
|---|---|---|---|---|
| The neuroinvasive potential of SARS-CoV2 may play a role in the respiratory failure of COVID-19 patients | J of Medical Virology | 27/2/20 | Review | 1 |
| Persistence of coronaviruses on inanimate surfaces and its inactivation with biocidal agents | J of Hospital Infection | 6/2/20 | Review | 2 |
| High temperature and high humidity reduce the transmission of COVID-19 | SSRN | 10/3/20 | Preprint | 3 |
| Identifying SARS-CoV-2 related coronaviruses in Malayan pangolins | Nature | 26/3/20 | Article | 4 |
| Early release - high contagiousness and rapid spread of severe acute respiratory syndrome coronavirus 2 - Volume 26, Number 7-July 2020 | Emerging Infectious Diseases | 7/4/20 | Research | 5 |

Although the top five articles for Dimensions were published in 2020, by 21 March they had all been cited at least 200 times in Dimensions (Figure 3), perhaps mainly by preprints, letters and short form fast publishing formats, such as brief communications, (academic) news, and case reports. All five documents exhibit a reasonably steady rate of increase. The simultaneous jumps in the lines presumably reflect weekly large scale database refreshing for Dimensions, although there were also smaller daily changes.

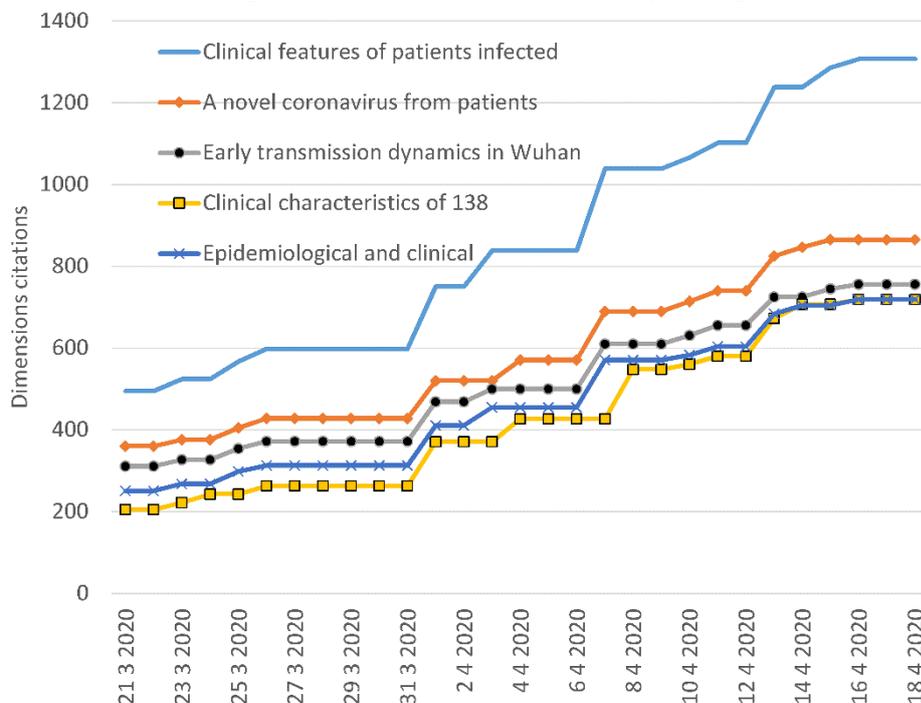

Figure 3. The cumulative number of Dimensions citations for the five most cited COVID-19 documents.

The top five Mendeley documents also started 21 March with a high number of readers, but almost five times more than the number of Dimensions citations (Figure 4). There was a similar pattern of steadily increasing numbers of Mendeley readers with periodic interruptions. In this case the interruptions resulted in temporary decreases in the numbers of Mendeley readers. This could be due to two factors. Either the database consolidates weekly, such as by merging duplicates, or its search is somehow weakened periodically so that the free text search (which is submitted in parallel with the DOI search) matches less documents. It is not possible to check which is correct from the data since Mendeley reports reader counts but not the identities of these readers.

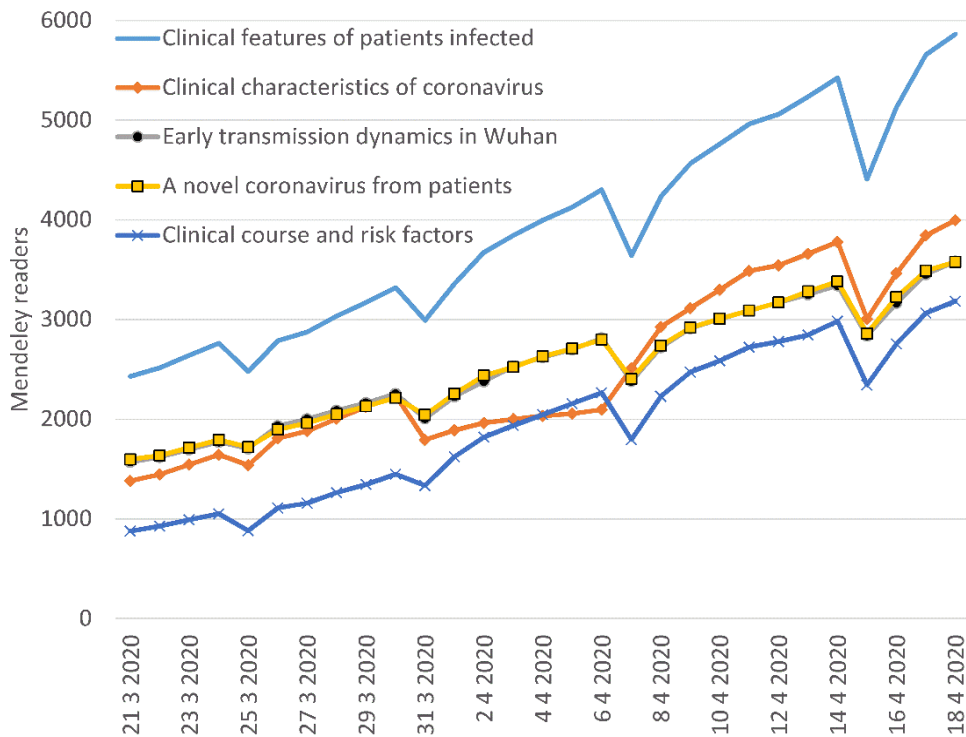

Figure 4. The cumulative number of Mendeley readers for the five most read COVID-19 documents.

Twitter shows a very different pattern to Dimensions and Mendeley. First, some of the documents are much younger, published during the date range analysed. Second, the number of tweeters achieves close to its maximum when first found by Dimensions, although this is not necessarily the original publication date.

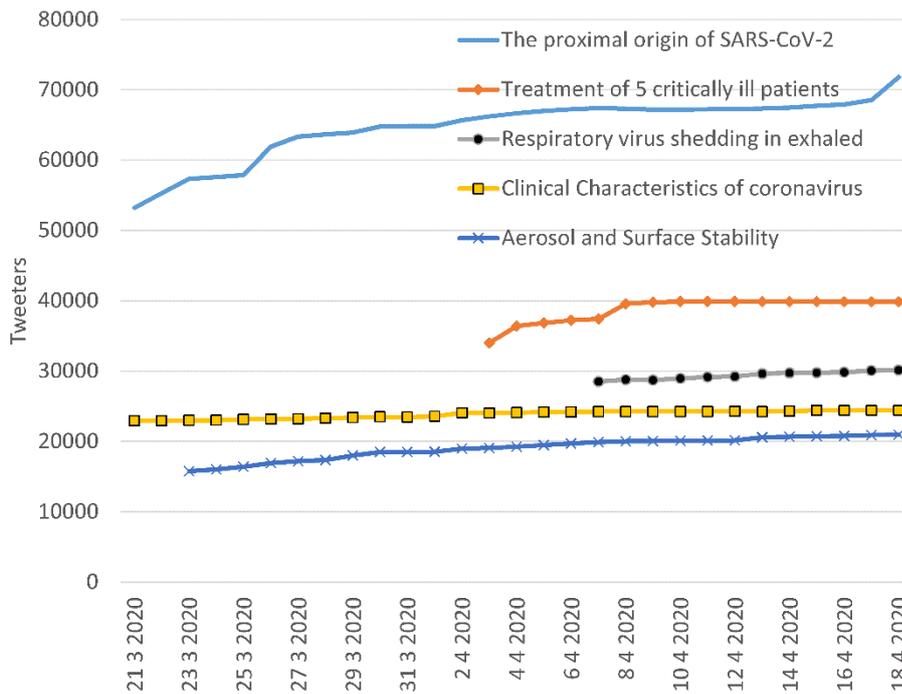

Figure 5. The cumulative number of Tweeters for the five most tweeted COVID-19 documents.

Facebook has a similar growth pattern to Twitter, except that there is a period of increasing interest for the proximal origin paper (Figure 6), which has a more moderate growth on Twitter. The (apparently speculative) news story about CERN scientists that was popular on Facebook did not get traction on Twitter and seems unlikely to be much cited or read.

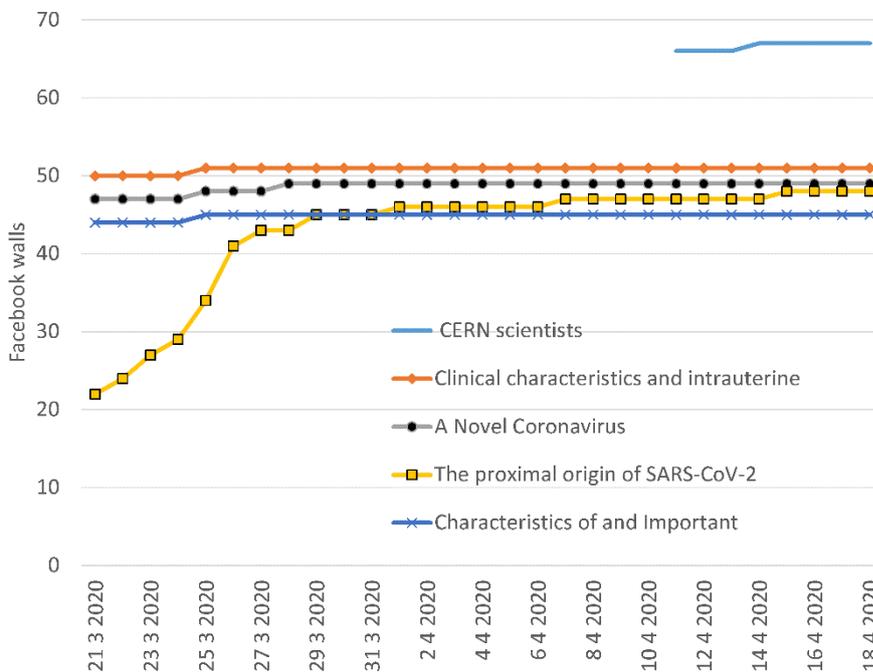

Figure 6. The cumulative number of Facebook wall posts for the five most walled COVID-19 documents.

The top news-cited articles were all covered by at least 400 news sources by the end of the period (Figure 7). Perhaps surprisingly, given that news is very time-dependant, all the sources experienced significant increases in the number of citing sources. Either Altmetric.com is constantly expanding its coverage of news sources (which is possible, but seems unlikely) or news stories about COVID-19 are prepared to cite old articles, perhaps for a more in-depth commentary or as background context for new articles.

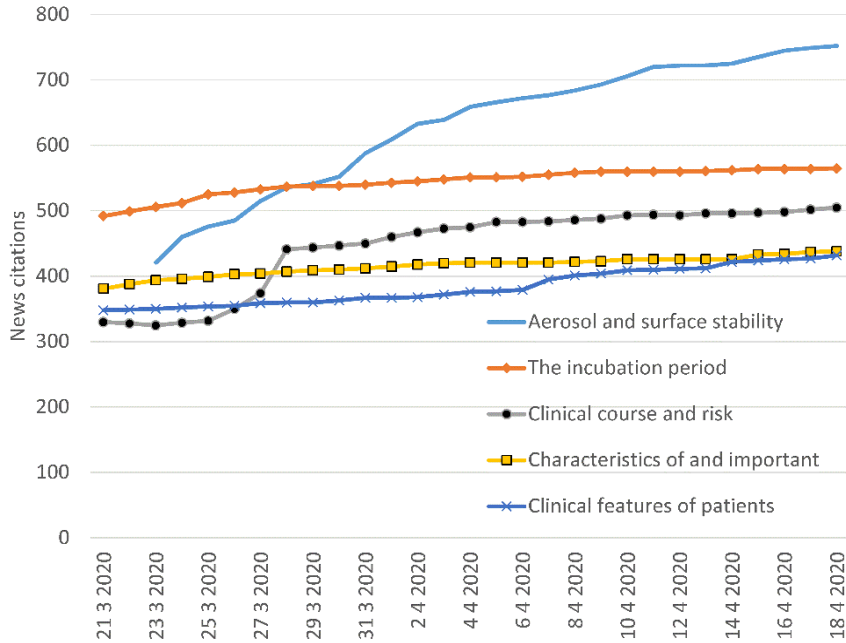

Figure 7. The number of news citations for the five most News-cited COVID-19 documents.

There were relatively few citations from Reddit, despite its use as a news source and many academic themes (subreddits) within the site (Figure 8). Perhaps reflecting its news status, older articles do not seem to increase their Reddit citation counts.

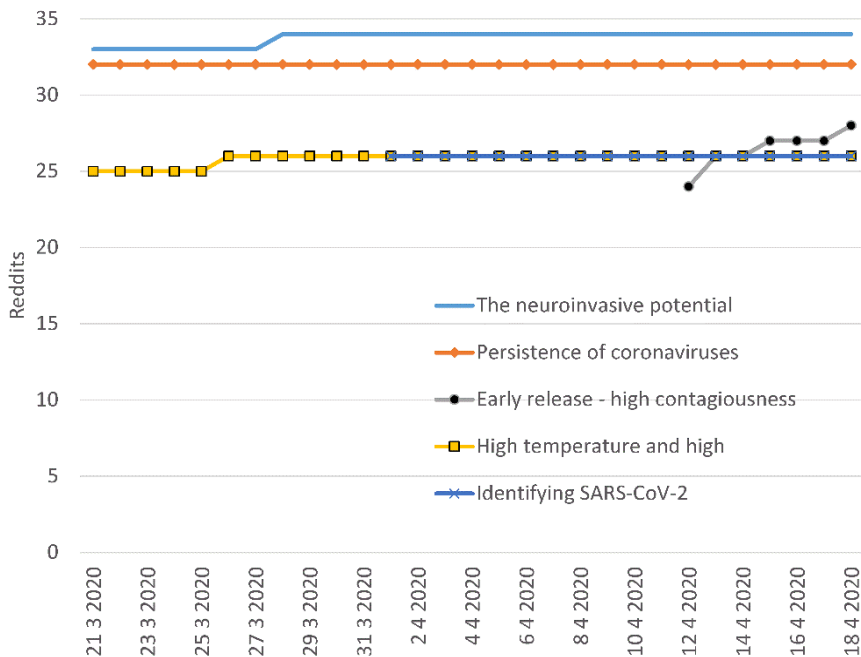

Figure 8. The cumulative number of Reddit posts for the five most posted COVID-19 documents.

*A comparison between average scores for different sources*

The 24th March was selected for a time series analysis because this date in the first week had the most new articles (349) found by Dimensions. For documents first found by Dimensions on 24 March 2020 and matching the COVID-19 queries, the average score was highest for Twitter and already above 1 on the start day (Figure 9). Average tweeter counts then increased slowly after the first few days. In contrast, average Mendeley reader counts for these 349 articles started close to zero and increased rapidly, except for weekly Mendeley indexing adjustments. Mendeley overtook Twitter after a week.

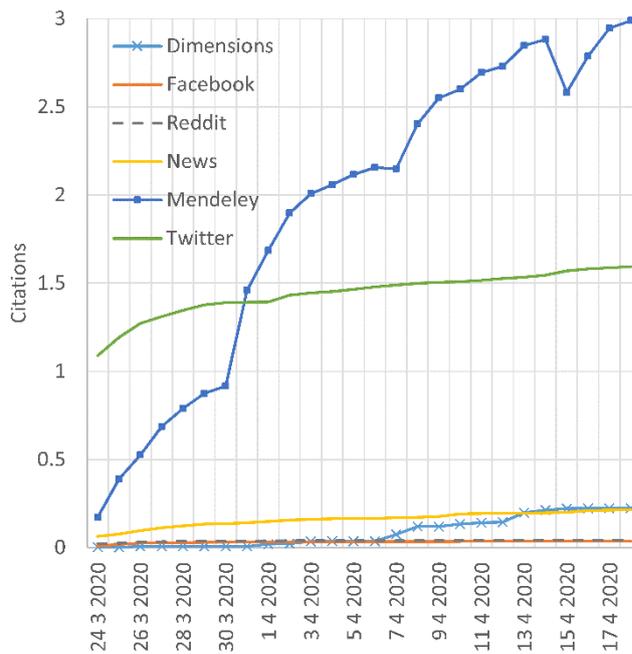

Figure 9. Daily average (geometric mean) citations by source for documents first found by Dimensions on 24 March 2020 (n=349).

The average citation counts for the remaining three sources were all much lower than for Mendeley and Twitter (Figure 10). Whilst Facebook and Reddit both displayed a similar growth pattern to Twitter (rapid initially, then slow), both News citations and Dimensions citations increased steadily. The average number of citations after three and half weeks is surprising for Dimensions, as is the constant growth for News sources.

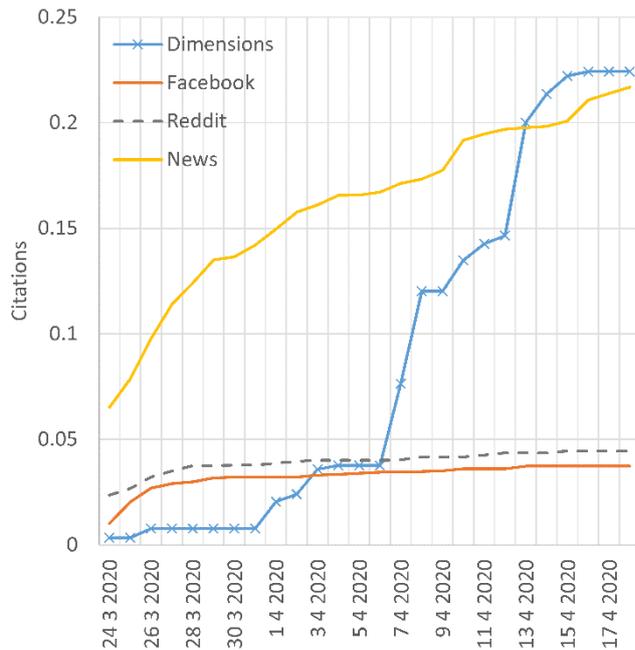

Figure 10. Daily average (geometric mean) citations by source for documents first found by Dimensions on 24 March 2020, excluding Mendeley and Twitter (n=349).

## Overlaps in citation counts between sources

Spearman correlation tests reveal the extent to which the same documents that are cited by one source are also cited by another source, together with the extent that they are cited. By April 18, 2020, correlations between Dimensions citations and altmetrics for documents first found by Dimensions in March 24 were strong, except for Reddit (Table 7). Since most (229; 66%) documents were uncited by 18 April, the correlation mainly confirms that, except for Reddit, news stories, publishing authors, and users of the different platforms tended to select the same documents for attention. The altmetrics also correlated moderately or strongly with each other, except for Reddit, in agreement with this conclusion. Thus, for the narrow topic of COVID-19, there seems to be a researcher-news-social media consensus about the most important topics, at least in the (very) short term.

The correlations (Table 8) do not take into account field differences or document type differences. The relatively high correlations could be at least partially due to ignoring contributions of low relevance to COVID-19, such as book chapters mentioning the possibility of a coronavirus 2, editorials, letters, and subject areas making relatively peripheral contributions to immediate needs.

Table 8. Spearman correlations between citation counts and altmetrics from 18 April 2020 for COVID-19 documents first found by Dimensions on 24 March 2020. All are statistically significant at p=0.001 (n=349).

|            | Mendeley | Twitter  | Facebook | News     | Reddit   |
|------------|----------|----------|----------|----------|----------|
| **Dimensions** | .653*** | .659*** | .453*** | .529*** | .249*** |
| **Mendeley**   | 1       | .689*** | .375*** | .473*** | .354*** |
| **Twitter**    |         | 1       | .411*** | .626*** | .363*** |
| **Facebook**   |         |         | 1       | .376*** | .251*** |
| **News**       |         |         |         | 1       | .335*** |

***Statistically significant at p=0.001

The influence of non-article document types on the correlations were tested by filtering out all non-articles. After manually removing documents that were not journal articles (mainly editorials, news, and letters), there were 106 standard journal articles (including reviews). Nevertheless, the correlations did not substantially change (Table 9). Some of the editorials were cited, explaining the lack of change. The positive correlations seemed to be due to articles with a stronger focus on COVID-19 being more noticed, whereas articles with a weaker focus on COVID-19 or giving weaker evidence were less noticed. Thus, both altmetrics and citations seem to focus on contributions that are more core to COVID-19 as a medical and public health issue.

The two most common Dimensions subject codes for the March 24 set were 1117 Public Health and Health Services (n=78) and 1103 Clinical Sciences (n=32). Except for Reddit (correlations close to 0), the pairwise correlations change little if the set is restricted to only subject categories 1117 or only 1103, with or without excluding non-article types. For example, the lowest correlation between Twitter and Dimensions for any of these four restricted sets is 0.638 (category 1103 with all document types, n=32). Thus, except for Reddit, the strong positive correlations between indicators do not seem to be due to field differences in the dataset.

Table 9. Spearman correlations between citation counts and altmetrics from 18 April 2020 for COVID-19 *journal articles* first found by Dimensions on 24 March 2020. (n=106).

|  | Mendeley | Twitter | Facebook | News | Reddit |
|---|---|---|---|---|---|
| **Dimensions** | 0.693*** | 0.734*** | 0.589*** | 0.585*** | 0.250** |
| **Mendeley** | 1 | 0.687*** | 0.401*** | 0.473*** | 0.316*** |
| **Twitter** |  | 1 | 0.562*** | 0.719*** | 0.382*** |
| **Facebook** |  |  | 1 | 0.440*** | 0.215* |
| **News** |  |  |  | 1 | 0.334*** |

*Statistically significant at p=0.05; **Statistically significant at p=0.01; ***Statistically significant at p=0.001

## Early altmetrics and later citation counts

Ideally, an indicator would help researchers and policy makers to identify important articles when they are first published, without having to wait for enough citations. To check for early evidence of later citation impact, the indicators were correlated with Dimensions citation counts on April 18, representing longer term citation counts (this is a weak proxy since decades are sometimes used for long term citations in other contexts, e.g., Stegehuis, Litvak, & Waltman, 2015).

On the day that a document is first findable in Dimensions, its tweeter count is the best indicator of likely long-term citation impact (Figure 11). Twitter users seem to be able to notice documents approximately on the date of first publication for their potential importance to COVID-19. After this date, the tweeter count does not increase much and its correlation with longer term Dimensions citations is stable. After about three weeks, Mendeley reader counts take over as a marginally better indicator of longer-term citation impact. It is not clear whether the same would be true for more mature citation counts, however, such as after a year. It is possible that early Dimensions citations (and Mendeley readers) reflect more temporary interest and are themselves highly influenced by the news or social sharing on Twitter, for example. The most cited sets of five papers analysed above suggest that highly recognised papers are particularly important for the disease, however. As above, this correlation ignores field differences and document type differences, although document differences seem to have little effect (Tables 8, 9).

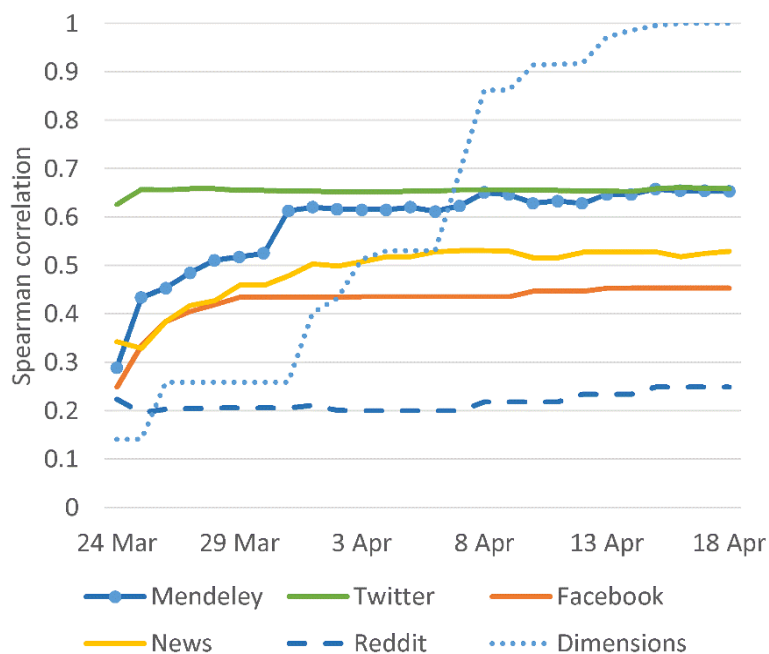

Figure 11. Spearman correlations between altmetrics and Dimensions citation counts from 18 April for COVID-19 documents first found by Dimensions on 24 March 2020 (n=349). Correlations with Dimensions citation counts on the same dates are also reported for context.

## Discussion

The results are limited by the range of factors mentioned in the Methods section. In particular, the coverage figures for the sources are not directly comparable due to the different scopes of the queries. In addition, the count data has not been field-normalized so the coverage comparisons do not reveal disciplinary differences. The correlations may also be exaggerated by not taking into account disciplinary differences. The results may show different patterns for earlier or later time periods. The properties of the scholarly databases and Altmetric.com's strategies may evolve over time, rendering the results obsolete. They may also not be applicable for later stages of COVID-19 research or for future epidemics or pandemics.

The COVID-19 query results comparison confirm the previous finding that COVID-related academic databases are appearing rapidly many different databases (Torres-Salinas, 2020). In addition, they confirm that Dimensions finds many publications not in Scopus and WoS but that Scopus indexes nearly all relevant publications found in the WoS core collection with the Conference Proceedings Citation Index. Presumably the difference would be smaller if other parts of WoS were included, such as the Book Citation Index, although the core collection includes the Emerging Sources Citation Index (Clarivate, 2020).

The results are not directly comparable to studies from before COVID-19 due to the unprecedented speed and volume of publishing on the topic. For example, Dimensions citation counts accrue more rapidly than previously reported for any topic. For comparison, the Scopus citations of 12 subject categories (full journal articles only) were a maximum of 0.12 in the month of publication, whereas the COVID-19 mixed set averaged almost double this after three weeks. The results are also qualitatively different in some respects. Whilst correlation tests have previously found tweeter counts to have little value as a scholarly impact indicator due to very low correlation with citation counts (Costas, Zahedi, & Wouters, 2015; Haustein, Larivière, Thelwall, Amyot, & Peters, 2014; Thelwall, Haustein, Larivière, & Sugimoto, 2013), the current study has found Tweet counts to be reasonable academic impact indicators and the best early impact indicator for the first three weeks. This may be partly due to the set of articles here covering multiple disciplines, but the results for the top-cited documents suggest that altmetrics are effective at pointing to the documents that are most central to COVID-19 as a medical and public health issue. Thus, the unprecedented threat of COVID-19 seems to have led to an unprecedentedly high and focused level of societal and academic attention being given to the most relevant research.

## Conclusions

The confirmed rapid increase in COVID-19 academic publications is encouraging in terms of the academic community rapidly reacting to the need for relevant research and commentaries. The importance of short form and quick contributions (viewpoints, correspondence, brief reports) is also evident in the highly cited papers, as is the importance of academic research for practical public health issues.

Despite the apparent high medical and public health value of some academic papers, the huge number of publications returned by a relevant search will presumably make the most important publications more difficult to find. This should not be a problem for medical researchers trained to use MeSH queries effectively, but might be problematic for other researchers, end users and the public, who may find bewilderingly many matches for their queries. The altmetric results suggest that altmetrics may be helpful for researchers needed to quickly identify the most useful new documents from the large number published daily. Altmetric counts may help to distinguish between core primary research and other contributions, such as editorial commentaries with narrower disciplinary or professional relevance (e.g., radiographers). Perhaps ironically, given that a core original goal for altmetrics was to develop indicators of societal impact that were different from scholarly impact indicators (Priem, Taraborelli, Groth, & Neylon, 2010), their greatest value (as early impact indicators) seems be occurring when the two concepts are most closely converging.